# Magnetization dynamics due to field interplay in field free spin Hall nano-oscillators

Ayush K Gupta[1], Sourabh Manna[2], Rajdeep Singh Rawat[2], Rohit Medwal[1*]

[1]Department of Physics, Indian Institute of Technology Kanpur, Kanpur 208016, India

[2]Natural Sciences and Science Education, National Institute of Education, Nanyang Technological University 637616, Singapore

[*]Correspondence should be addressed to Rohit Medwal: rmedwal@iitk.ac.in

**Abstract:** Spin Hall nano oscillators (SHNOs) have shown applications in unconventional computing schemes and broadband frequency generation in the presence of applied external magnetic field. However, under field-free conditions, the oscillation characteristics of SHNOs display a significant dependence on the effective field, which can be tuned by adjusting the constriction width, thereby presenting an intriguing area of study. Here we study the effect of nano constriction width on the magnetization dynamics in anisotropy assisted field free SHNOs. In uniaxial anisotropy-based field free SHNOs, either the anisotropy field ($B_{anis}$) or the demagnetization field ($B_{demag}$) can dominate the magnetization dynamics depending on the constriction width. Our findings reveal distinct auto-oscillation characteristics in narrower constrictions with 20 nm and 30 nm constriction width compared to their wider counterpart with 100 nm width. The observed frequency shift variations with input current ($I_{DC}$) and constriction widths stem from the inherent nonlinearity of the system. The interplay between the $B_{demag}$ and $B_{anis}$, coupled with changes in constriction width, yields rich dynamics and offers control over frequency tunability, auto oscillation amplitude, and threshold current. Notably, the spatial configuration of spin wave wells within the constriction undergoes transformations in response to changes in both constriction width and anisotropy. The findings highlight the significant influence of competing fields at the constriction on the field-free auto oscillations of SHNOs, with this impact intensifying as the constriction width is varied.

# I. INTRODUCTION

Our dependence on technological devices has been increasing, and we are striving to find faster and more efficient ways to compute and communicate. Current complementary metal-oxide semiconductors (CMOS) technology [1] used in a wide range of devices may not be able to keep up with increase in novel applications of computing devices [2,3]. Spin-based devices such as Spin torque magnetic random-access memory (STMRAM) [4], Spin torque nano oscillator (STNO) [5], Spin Hall nano oscillator (SHNO) can be used to enhance our current computing process [6] and data storage capabilities [7]. Spin Hall nano oscillators have been of particular interest in recent times due to their easy fabrication process, scalability, and CMOS compatibility. In addition, a wide variety of spin-wave modes, such as propagating SW [8–10], droplet solitons [11,12], localized bullet modes [13,14] etc. can be excited by exploiting the inherent non-linear magnetization dynamics in SHNOs. This nonlinearity of the magnetization dynamics can be well tuned via input dc current as well as gate voltage [15,16], that enables novel applications of SHNO in magnonics [8,17,18]. In summary, SHNOs offer a plethora of applications as broadband frequency generation [19,20], high frequency tunability [20,21], mutual synchronization [21–24], and various non-conventional computing schemes [25–27],.

SHNO devices utilize the spin Hall effect (SHE) [28] to generate high-frequency oscillations. This phenomenon occurs when a charge current flows through a material with significant spin-orbit coupling (SOC), such as heavy metals. In such a high SOC material, the charge current segregates based on the spin polarization of electrons, leading to the production of a pure spin current transverse to the charge current. SHNOs are typically bilayer heterostructures which consist of a ferromagnet layer (FM) attached to a heavy metal layer (HM). Passing a charge current through the HM layer results in a pure spin current [29] due to SHE and subsequent injection of spin current into the FM layer. The injected spin current exerts a damping like spin orbit torque (SOT) [30] on the magnetization that may counter the damping torque depending on the orientation of magnetization with respect to the spin polarization vector.

Therefore, a sufficiently strong damping like SOT acting opposite to the damping torque can fully compensate for the damping and lead to a sustained oscillation of magnetization around the effective field $H_{eff}$. Typically, a biasing magnetic field is used for the operation of SHNO. However, such a biasing field is not the indispensable factor for the generation of sustained auto-oscillation in SHNO as demonstrated in a few recent studies of field-free SHNO [31–33]. Designing field-free SHNO will provide additional benefits of lower energy consumption, higher scalability, better scope for CMOS integration etc.

Awad et al [34] has experimentally demonstrated the importance of constriction width in a nanoconstriction SHNO to determine the oscillation characteristics in presence of a sufficiently high out-of-plane biasing field. In this article we present a systematic study on the effect of constriction width on the field free magnetization oscillations considering the in plane uniaxial anisotropy in the FM layer. Therefore, $H_{eff}$ in this case has contribution from the anisotropy field ($B_{anis}$) in addition to the demagnetisation field ($B_{demag}$) due to dipolar interactions and the current induced Oersted field. The demagnetisation field is highly dependent on the constriction width; decreasing the width brings the edges of constriction closer which in turn increases the dipolar interaction energy as well as the $B_{demag}$. On the other hand, $B_{anis}$ is invariant of constriction width and instead depends on the magnetocrystalline anisotropy in the system. Consequently, wider ($B_{anis} > B_{demag}$) and narrow constrictions ($B_{anis} < B_{demag}$) have different fields as dominating component in total internal field $B_{int}$, at the constriction. In the later part of this paper, we show that frequency response for the different constriction widths is also a subject of the interplay between the anisotropy field and the demagnetisation field. We further show that the nonlinearity coefficient $N$, which describes the inherent nonlinearity in frequency of the auto-oscillating modes, is a function of $B_{int}$ and changes sign with decrease in constriction width. In the final part, we discuss about the substantial difference of the spatial profile of auto oscillation modes at varying constriction width.

## II. METHODOLOGY

The magnetization dynamics of a ferromagnet is governed by two major torques−a precessional torque due to the effective magnetic field and a damping torque that tries to damp the precession about the effective field ($H_{eff}$) due to the electron-phonon, electron-impurities scattering events in the material. The magnetization dynamics can be expressed by the Landau-Lifshitz-Gilbert (LLG) equation given as.

$$\frac{\partial M}{\partial t} = (-\gamma M \times B_{eff}) + \left(\frac{\alpha}{M_s} M \times \frac{\partial M}{\partial t}\right) \quad (1)$$

Here the first term on the RHS is the precessional torque $\tau_p$ acting on the magnetization, which is tangential to the precession trajectory and the second term is the damping torque $\tau_D$ pointing towards the $B_{eff}$. The $B_{eff} = B_{ext} + B_{demag} + B_{anis}$, consists of the external field ($B_{ext}$), demagnetization field ($B_{demag}$) and anisotropy field ($B_{anis}$) present in the material.

To obtain sustained precession of magnetization around the $B_{eff}$, a so-called anti-damping torque opposite to the direction of $\tau_D$ must be present in the system to counter the damping. This additional anti-damping torque can be obtained from the injected spin current. The high spin-orbit coupling in the HM layer facilitates the conversion of the applied charge current into a transverse spin current denoted by $J_s = \left(\frac{I_{DC}}{I_{ref}}\right)\theta_{SH}|J_c|$ [11]. Here, $\theta_{SH}$ is the spin Hall angle of HM, $|J_c|$ is the current density at $I_{ref}$. The injected $J_s$ gives rise to the anti-damping torque $\tau_{AD}$ that acts opposite to the damping torque and tries to orient the $M$ along the spin polarization direction $\boldsymbol{\sigma}$. The $\boldsymbol{\tau_{AD}}$ is typially of Slonczewski form which is given by [35].

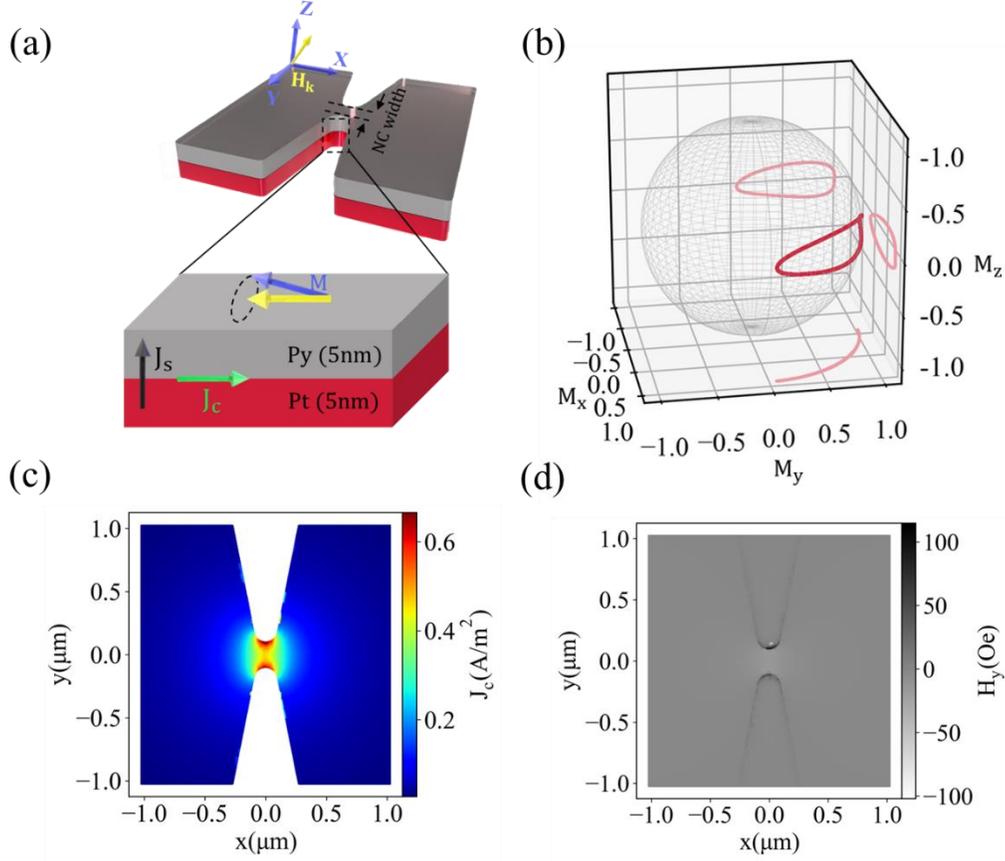

*Figure 1. (a) System schematic shows the SHNO and directions of spin current, charge current and $H_k$. (b) Trajectory traced by head of magnetization for 100 nm constriction for $Ku_1 = 10\frac{kJ}{m^3}$ at $I_{th}$ along with projections. (c) Spatial profile of current density for 100 nm constriction width. (d) Spatial profile of y component of Oersted field for 100 nm constriction width.*

$$\tau_{AD} = \frac{|J_c|\hbar\theta_{SH}}{2et_{FM}\mu_o M_s}(\boldsymbol{M} \times (\boldsymbol{\sigma} \times \boldsymbol{M})) \qquad (2)$$

Here $t_{FM}$ is the thickness of the FM layer. Field-like torque $\boldsymbol{\tau_{FL}}$ due to the $\boldsymbol{J_s}$, which is analogous to the precessional torque, is not considered due to its low magnitude as compared to the $\boldsymbol{\tau_{AD}}$ [36]. The magnetization dynamics in this case is governed by the Landau Lifshitz Gilbert Slonczewski (LLGS) equation [30,37].

$$\frac{\partial \mathbf{M}}{\partial t} = (-\gamma \mathbf{M} \times \mathbf{H}_{eff}) + \left(\frac{\alpha}{M_s} \mathbf{M} \times \frac{\partial \mathbf{M}}{\partial t}\right) + \frac{|J_c|\hbar \theta_{SH}}{2et_{FM}\mu_o M_s}(\mathbf{M} \times (\boldsymbol{\sigma} \times \mathbf{M})) \quad (3)$$

The spin polarization efficiency $\epsilon = \frac{\theta_{SH}\Lambda^2}{(\Lambda^2+1)+(\Lambda^2-1)(M.\sigma)}$ describes the transparency of the interface for spin current [38], where $\Lambda$ describes the transport properties of the material depending on the angle between $\mathbf{M}$ and $\boldsymbol{\sigma}$. For pure spin current based on spin orbit torque $\Lambda = 1$, thus $\epsilon = \frac{\theta_{SH}}{2}$ [39].

To examine the width dependent field free auto-oscillations, we have designed the SHNO as a Py (5 nm)/Pt (5 nm) bilayer stack with nano constriction geometry and varying constriction widths. The stack, measuring $5\,\mu m \times 5\,\mu m$, incorporates a round nano constriction at the center, characterized by an opening angle of 22° and a radius of curvature of 50 nm. Using COMSOL®, this geometry is simulated in air, with an electric current applied to the yz face (Fig. 1(a)) along x-direction. The current density distribution and the associated Oersted field profile (Fig. 1(b) and (c)) are obtained from COMSOL® simulations, at the reference current $I_{ref} = 1\,mA$. Given the relatively lower conductivity of the FM layer ($3.12 \times 10^6\,S/m$) compared to the HM layer ($8.9 \times 10^6\,S/m$), the majority (74%) of the current flows through the HM layer, rendering the spin filtering effect negligible in the FM layer. The SHE in Pt leads to the generation of pure spin current with $-y$ polarization that diffuses into the Py layer, perpendicular to the xy plane. The Py layer is discretized into a grid of $512 \times 512 \times 1$ rectangular cells, with each cell measuring $2 \times 2 \times 5\,nm^3$, which is smaller than the Py's exchange length along the lateral directions. The material parameters of Py are defined as follows: gyromagnetic ratio $\gamma = 29.53\,GHz/T$, damping constant $\alpha = 0.02$, saturation magnetization $M_s = 8 \times 10^5\,A/m$, and exchange constant $A_{ex} = 12 \times 10^{-12}\,J/m$ [8]. The magnetization dynamics is obtained by numerically integrating the LLG equation. Initially, the magnetization is relaxed for 5 ns before applying $I_{DC}$. The magnetization dynamics is simulated for a total of 45 ns for each $I_{DC}$ value, excluding the initial 15 ns to account for stable precession and eliminate transient effects.

To eliminate the requirement for an external biasing field, we exploit the inherent uniaxial anisotropy of the ferromagnetic material. The uniaxial anisotropy generates the anisotropy field, which, combined with the demagnetization field and Oersted field, forms the effective field ($\boldsymbol{H_{eff}}$) responsible for the precession of magnetization. In order to investigate the oscillation properties of such a bias-free SHNO under the influence of an anisotropic field, we performed the simulation with varying the DC current, uniaxial anisotropy constant ($K_u$) and the orientation of easy axis. Easy axis angle is defined such that it maximizes the amplitude of auto-oscillations. This angle approaches the y-axis for narrower constrictions, while increasing the constriction width causes the angle to shift towards the x-axis (see Appendix C (Fig. 8)). This behaviour arises from the dipolar interaction originated from the geometrical confinement at the constriction region which is characterized by the constriction width. (See Table 1).

## III. RESULTS

The width dependent field free auto oscillations are studied for NC width ranging from 100 nm to 20 nm at uniaxial anisotropy values varying from $K_u = 5\ kJ/m^3$ to $15\ kJ/m^3$. The magnetization trajectory during auto oscillations for 100 nm at $K_u = 10\ kJ/m^3$ is shown in Fig 1(b). It was observed that the dynamics for 100 nm show a red shift in frequency with an increase in current, regardless of the anisotropy values. Increasing the anisotropy ($K_u$) led to an amplitude increase due to the closer proximity of spin wave wells which is caused by an elevated $\boldsymbol{B_{int}}$. The red shift in frequency was observed till the constriction width of 40 nm (see Appendix E), where all the constriction had a negative $\frac{\delta f}{\delta I}$. Conversely, for 30 nm constriction, initially, a blue shift in frequency is observed with almost no frequency tunability at high $I_{DC}$ (Fig. 2).

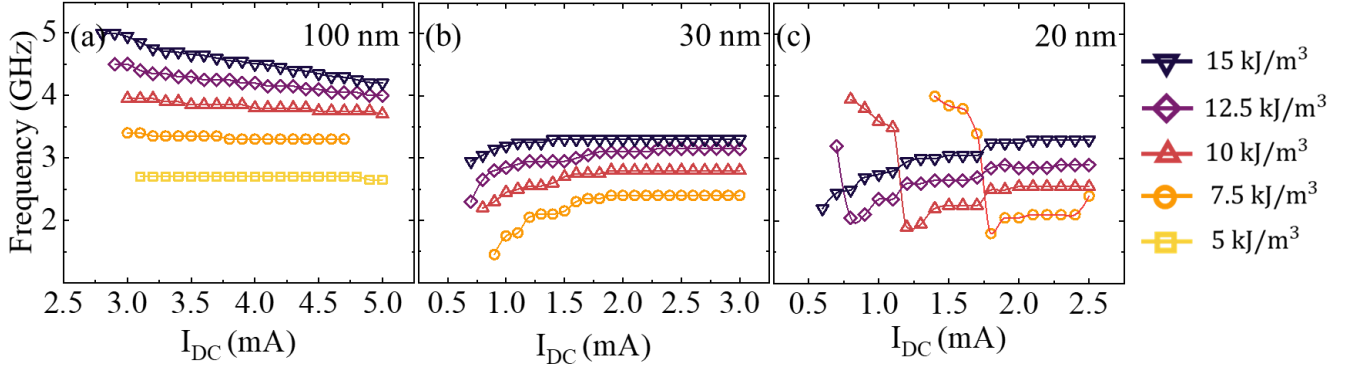

***Figure 2.*** *Variation of the frequency of auto oscillations with input current for different constriction sizes at different $K_u$ values.*

The spin wave (SW) wells for these narrow widths are mostly restricted to constriction. On the flip side, the frequency response of the 20 nm constriction exhibited a consistent blue shift with current, accompanied by significantly more constrained spin wave wells. These different magnetodynamical behaviour originate from the change in the total internal magnetic field $\boldsymbol{B_{int}}$ at the constriction. In our case the $\boldsymbol{B_{int}}$ also has an additional contribution from $\boldsymbol{B_{anis}}$ along with $\boldsymbol{B_{Oe}}$, $\boldsymbol{B_{demag}}$ due to the uniaxial anisotropy present in the system. The anisotropy field $\boldsymbol{B_{anis}}$ is represented by.

$$\boldsymbol{B_{anis}} = \frac{2K_u}{M_s} \cos\theta \, \hat{u} \qquad (4)$$

The $\theta$ is the angle between magnetization **M** and the easy axis and $\hat{u}$ is the direction of the easy axis. The $\boldsymbol{B_{anis}}$ decreases as the amplitude of the oscillations is increased. The spin torque $\boldsymbol{\tau_s} \propto J_c$ [30], as a consequence the amplitude scales with an increase in current which in turn reduces the $\boldsymbol{B_{anis}}$. The dependence of $\boldsymbol{B_{demag}}$ on auto oscillation amplitude is attributed to the magnitude of $\boldsymbol{B_{demag}}$ being proportionate to the component of magnetization along the corresponding direction. Increase in auto oscillation amplitude decreases the component of magnetization along the easy axis, leading to a diminished contribution from $B_{demag}$. Now the $B_{int}$, taking $B_{anis}$ and $B_{demag}$ into account, can be written as

$$B_{int} = \frac{2K_u}{M_s} cos\theta u_i m_i - B^i_{demag} \qquad (5)$$

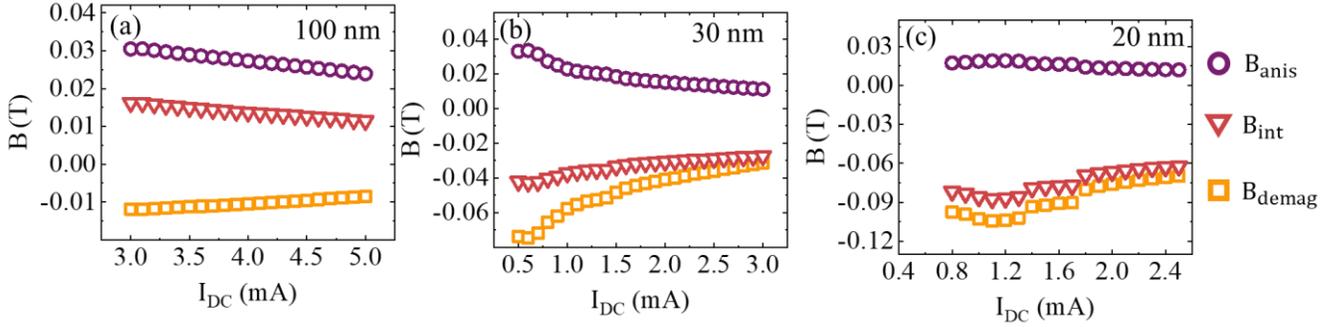

*Figure 3. Variation of the y-component of $B_{anis}$, $B_{demag}$ and $B_{int}$ at the centre of constriction with input current for different constriction width at $K_u = 10 \ kJ/m^3$.*

Fig. 3 illustrates the current dependence of $B_{anis}$, $B_{int}$ and $B_{demag}$ at $K_u = 10 \ kJ/m^3$ for different constriction widths. At all three constrictions, a decrease in $B_{anis}$ and $B_{demag}$ is observed with an increase in current. This is due to the increased amplitude of oscillation with the current. It is also interesting to see that for the smallest width of 20 nm, $B_{demag}$ is ten times larger than that of the 100 nm, which can be explained by the fact that the dipolar interaction energy at the center of the constriction varies inversely with the constriction width.

The major contribution to $B_{int}$ for 20 nm NC comes from the $B_{demag}$ (Fig. 3(c)) while for 100 nm NC width the predominant contribution comes from $B_{anis}$ (Fig. 3(a)). Consequently, the $B_{int}$ for small NC width follows $B_{demag}$ instead of $B_{anis}$, contrary to the behaviour observed in large constriction widths (see Appendix A Fig. 6).

The frequency tunability of the auto oscillations is a crucial aspect of the SHNO. The frequency of auto-oscillations $f$ is extracted by performing FFT over the stable regime of oscillations. Current tunability of $f$ with varying constriction width and $K_u$ is shown in Fig. 2. Notably, the magnetodynamical behaviour of field free SHNOs is strikingly different for all three constriction widths. This can be described by the magnetodynamic nonlinearity of the system. In the case of a thin film with in-plane fields, the complex spin wave amplitude is described by [14]

$$\frac{\delta b}{\delta t} = -i[\omega_o - D\Delta b + N|b|^2 b] - \Gamma b + f\left(\frac{r}{R_c}\right)\sigma I b - f\left(\frac{r}{R_c}\right)\sigma I |b|^2 b \tag{6}$$

Here, $\omega_o$ is the FMR frequency, $D = \frac{2A_{ex}}{M_s}\frac{\delta\omega_o}{\delta H}$ is spin wave dispersion coefficient, $R_c$ is related to the effective area of spin current injection, $\Gamma = \alpha\left(\omega_H + \frac{\omega_M}{2}\right)$ is spin wave dissipation rate with $\omega_H = \gamma H_{int}, \omega_M = 4\pi\gamma M_s$. $N$ is the non-linear frequency shift given by.

$$N = -\frac{\omega_H \omega_M \left(\omega_H + \frac{\omega_M}{4}\right)}{\omega_o\left(\omega_H + \frac{\omega_M}{2}\right)} \tag{7}$$

Considering a standing spin wave bullet solution, $b = B_o \psi\left(\frac{r}{l}\right) e^{-i\omega t}$. Where $\psi$ describes the spatial profile of the bullet. The frequency of the oscillations is given as $\omega = \omega_o + NP$. Where $P = |B_o|^2$ is the power of the oscillations, which directly relates to the amplitude. The value of $N$ depends on the $\boldsymbol{B_{int}}$ and differs for all three constrictions. For SHNOs with in plane fields the value of N can be approximated with equation (7) (see Appendix D). It can be clearly deduced from the expression of $N$, that for positive $\boldsymbol{B_{int}}$, the non-linear frequency shift will be negative. This is evident in the case of the 100 nm constriction (Fig. 3(a)), where $\boldsymbol{B_{int}}$ is dominated by $\boldsymbol{B_{anis}}$ and stays positive (see Appendix A (Fig. 6)) for all $K_u$ values throughout the $I_{DC}$ sweep. Consequently, the frequency of the oscillations for this constriction decreases with current, exhibiting a red shift. In contrast, the value of N for 30 nm constriction is positive but stagnates for high $I_{DC}$ because of minuscule change in $\boldsymbol{B_{int}}$. The frequency response shows a blue shift for low $I_{DC}$, for high $I_{DC}$ there is little to no frequency tunability with current (Fig. 2(b)). The frequency response for 20 nm constriction shows a blue shift for about all current values due to the positive value of non-linear frequency shift. (Fig. 2(c)). An increase in frequency with $K_u$ is also observed which is attributed to that the frequency is characterized by the $B_{int}$. Increment in $K_u$ increases the $B_{anis}$'s contribution to $B_{int}$, which increases the frequency.

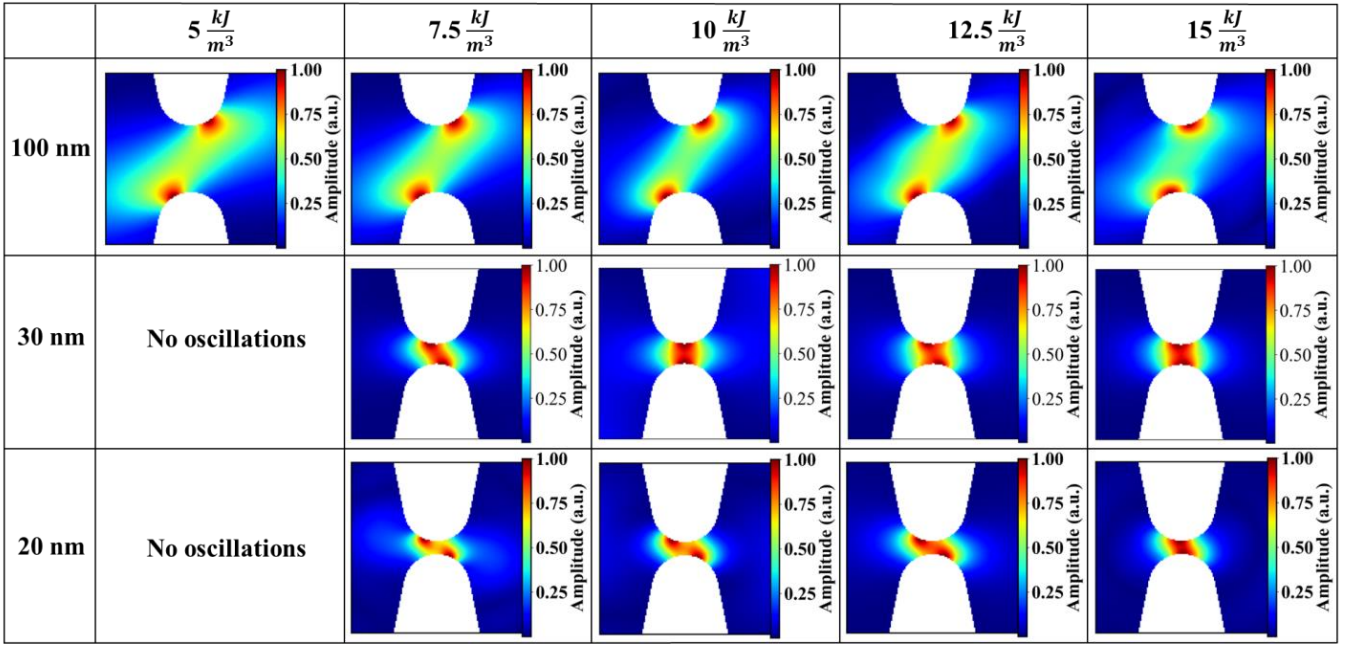

*Figure 4. Spatial profile of auto oscillations at threshold current density for different constriction widths and $K_u$ values.*

The threshold current density for auto oscillations increases with decrease in $K_u$ and constriction width. The $I_{th}$ depends on the $\boldsymbol{B_{int}}$, from the macrospin model $I_{th}$ can be written as [40]

$$I_{th} = \frac{\gamma \alpha}{2|\tau_{DL}|(\hat{m}_o \cdot \hat{\sigma})}\left(\mu_o M_s + 2B_{int} - \frac{B_{int}^2}{2M_s}\right) \tag{8}$$

Here, $|\tau_{DL}|$ is the magnitude of damping like SOT, $(\hat{m}_o.\hat{\sigma})$ is the angle between relaxed magnetization and spin polarization. According to the equation above, the threshold current $I_{th}$ should decrease with increasing $B_{int}$. Fig. 2(c) shows a sharp increase in $I_{th}$ with decreasing $K_u$ for 20 nm constriction as the change in $\boldsymbol{B_{int}}$ is very steep at low constriction widths. For 20 nm and 30 nm constriction, the auto oscillation could not be excited for $K_u = 5\ kJ/m^3$ even at current densities of approximately $7\ TA/m^2$. This limitation arises from the substantial $\boldsymbol{B_{demag}}$ at the centre of the constriction, which hinders oscillation excitation as the SOT fails to overcome $\tau_D$.

A spatial profile of the auto oscillation amplitude is shown in Fig. 4, which is obtained using the FFT of spatiotemporal data of the oscillations in each cell. Notably, the localization of amplitude varies across different constriction widths. The formation of spin wave (SW) wells at the constrictions follows the distribution of $\boldsymbol{B_{int}}$. In sufficiently wide constrictions, as the current increases, SW modes emerge at opposite edges of the constriction, gradually approaching each other with increase in $K_u$. Beyond a critical current, these modes detach from the edges and begin propagating. These SW wells move closer to each other and try to align with the easy axis as the increase in $K_u$ is results in the enhancement in $\boldsymbol{B_{int}}$. The alignment of edge modes at the opposite constriction edges in narrow constrictions differs from that of 100 nm constrictions due to the proximity of the constriction edges. This proximity leads to the dominance of the demagnetization field over other fields, resulting in a negative sign for the $B_{int}$. The opposite direction of the $B_{int}$ at different constrictions, results in SW wells having positive and negative slope from x-axis respectively for wider and narrower constrictions. Additionally, in 20 nm and 30 nm constrictions, the spin wave (SW) wells occupy the entire available volume within the constriction.

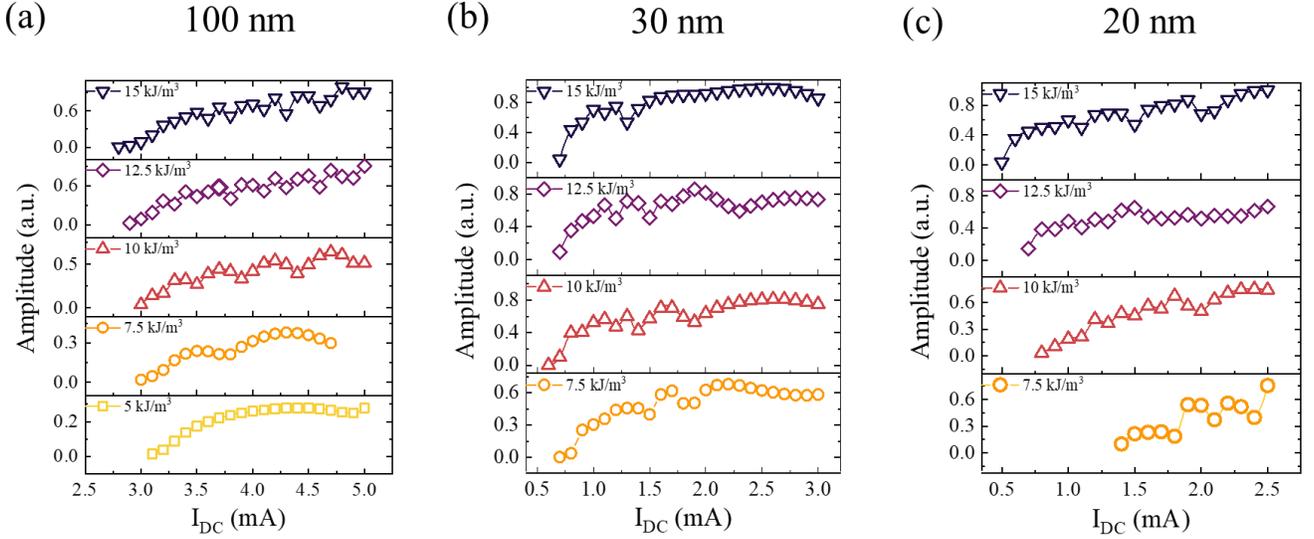

***Figure 5.*** *Variation of amplitude of auto oscillations with input current for different constriction sizes at different $K_u$ values. (a) 100 nm (b) 30 nm (c) 20 nm.*

An increment in amplitude of the auto oscillations with uniaxial anisotropy and current for all the constriction widths is observed (Fig. 5). Higher amplitude at higher current densities is due to the increased SOT at the constriction, which pulls the magnetization far out of its equilibrium. To understand the variation of amplitude with the $K_u$ the $\boldsymbol{B_{int}}$ distribution at the constriction for all three constrictions is examined. The change in $\boldsymbol{B_{int}}$ at a similar current density for different $K_u$ is due to the change in the $\boldsymbol{B_{anis}}$ which competes with the $\boldsymbol{B_{demag}}$ as both are opposite, thus the high $\boldsymbol{B_{anis}}$ results in a high $\boldsymbol{B_{int}}$. The SW wells move closer to each other due to the high $\boldsymbol{B_{int}}$ resulting in a larger auto-oscillation area at the constriction, which increases the auto-oscillation amplitude**.** Additionally, a decrease in auto oscillation amplitude with constriction width is observed. This is attributed to very high $\boldsymbol{B_{int}}$ amplitude due to the dominating $\boldsymbol{B_{demag}}$. High $|B_{int}|$ gives rise to strong damping torque which reduces the auto oscillation amplitude.

The fluctuations in the amplitude are also observed, because at high $K_u$ values the amplitude of the oscillations is shared between multiple harmonics. When the minor harmonics have high amplitude, the amplitude of the major harmonic shrinks. This behaviour is also visible in the linewidth spectra of the auto oscillations (see Appendix B). The low linewidth corresponds to the high amplitude of the major harmonic; on the other hand, high linewidth appears due to the significant amplitude of other minor harmonics.

## IV. CONCLUSION

In summary, we have demonstrated that the uniaxial anisotropy-based field free SHNOs exhibit significant variations in auto oscillation properties with change in constriction width. Wide constrictions are primarily influenced by the anisotropy field, resulting in a consistently negative frequency shift with current. In contrast, narrow constrictions are predominantly governed by dipolar interactions, leading to a primarily positive frequency shift. Decreasing the constriction width results in a lower oscillation frequency due to the increased central magnetic field $B_{int}$ at the centre of the constriction. Additionally, the oscillation of SW modes forming at the constriction have different behaviour with changes in constriction width. For 100 nm constriction the SW wells are aligned at the opposite edges of the constriction with a positive slope from x-axis, while for the narrow constrictions where $B_{demag}$ dominates the SW wells have a negative slope. Furthermore, the SW modes for wide constrictions stick to the edges of the constriction whereas for narrower (20 and 30 nm) constrictions these modes are inherently different and predominantly occupy the entire constriction region from the start. Our results show that the miniaturization of field-free SHNOs essentially leads to lower threshold current of auto oscillation. However, wider constriction provides higher oscillation amplitude and better frequency tunability with input current. These findings illuminate the field-free auto-oscillation dynamics of SHNOs across varying length scales, opening avenues for their application in diverse unconventional computing schemes like neuromorphic computing, reservoir computing, and probabilistic computing.

**Data availability statement**

The data that support the findings of this study are available from the corresponding author upon reasonable request.

**Table I**. Easy axis angle and maximum value of $J_c$ at constriction for different constriction widths.

| NC width (nm) | $J_c \left(\frac{TA}{m^2}\right)$ | $\phi$ (degrees) |
|---|---|---|
| 20 | 2.727 | 90 |
| 30 | 1.818 | 90 |
| 40 | 1.407 | 80 |
| 60 | 1.006 | 80 |
| 80 | 0.797 | 70 |
| 100 | 0.665 | 70 |

# APPENDIX A: $B_{int}$ FOR DIFFERENT CONSTRICTION WIDTHS AT DIFFERENT $K_u$ VALUES

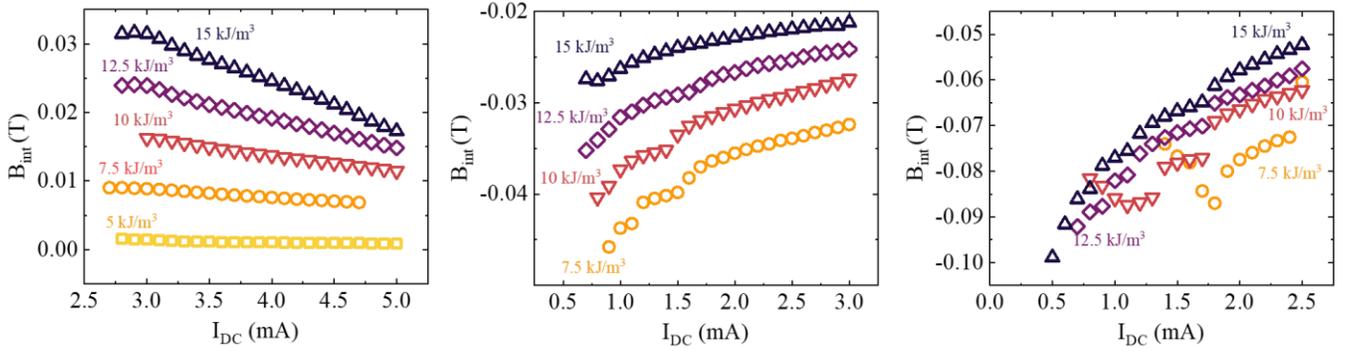

**Figure 6**: $B_{int}$ at constriction widths at different $K_u$ as a function of current have been plotted (a) 100 nm (b) 30 nm (c) 20 nm.

The plot displays $B_{int}$ for various constriction widths and $K_u$ values. In the case of the 100 nm width, $B_{anis}$ dominates, leading to an increase in $B_{int}$ with higher $K_u$. Additionally, at larger $K_u$ values, $B_{int}$ steeply decreases due to the increased auto oscillation amplitude. For narrower constrictions, $B_{int}$ becomes negative as a result of the heightened influence of $B_{demag}$. At higher $K_u$ values, $B_{int}$ becomes less negative for narrower constrictions and exhibits a higher positive value for wider constriction, primarily due to the increasing contribution from $B_{anis}$.

# APPENDIX B: LINEWIDTH VARIATION OF AUTO OSCILLATIONS WITH CONSTRICTION WIDTH, INPUT CURRENT AND $K_u$

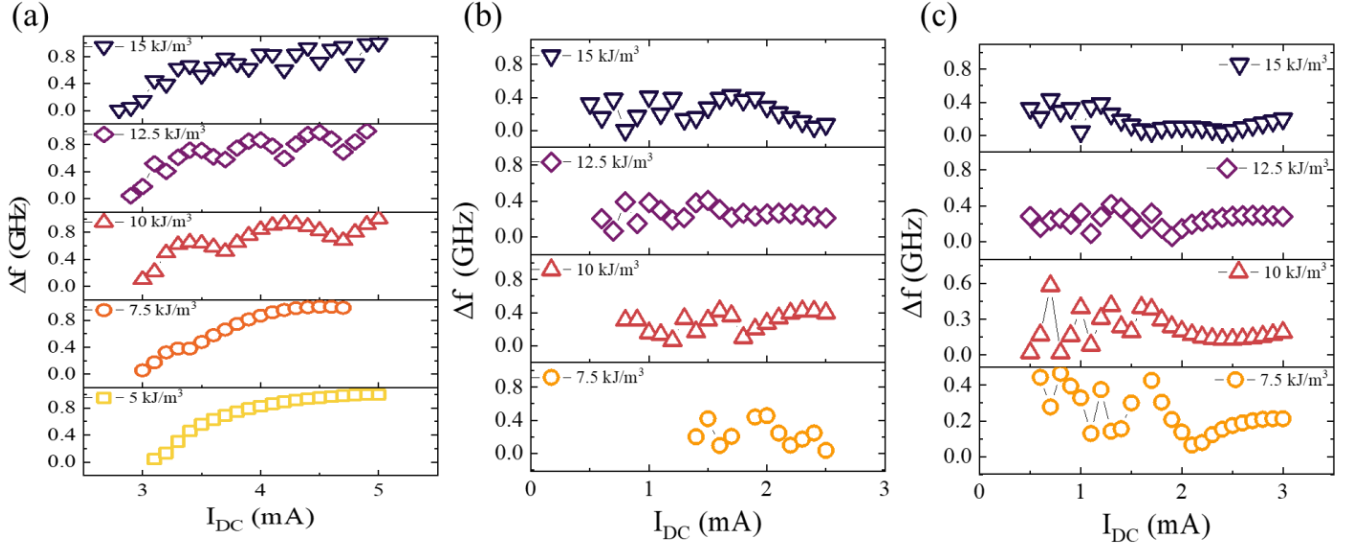

**Figure 7**: *Linewidth of auto oscillations at different $K_u$ as a function of current have been plotted (a) 100 nm (b) 20 nm (c) 30 nm.*

For the 100 nm constriction, the fluctuations in Δf increase as $K_u$ is raised. The oscillation amplitude undergoes harmonic splitting due to high $B_{int}$ at higher $K_u$ values, resulting in broadened peaks. In contrast, for the 20 nm and 30 nm constrictions, Δf fluctuations occur at lower $I_{DC}$ values but diminish as $B_{int}$ stabilizes. Lower constrictions exhibit improved linewidth, potentially attributable to the proximity of edge modes within the constriction.

# APPENDIX C: ANGULAR VARIATION OF AUTOSCILLATION AMPLITUDE FOR DIFFERENT CONSTRICTION WIDTHS

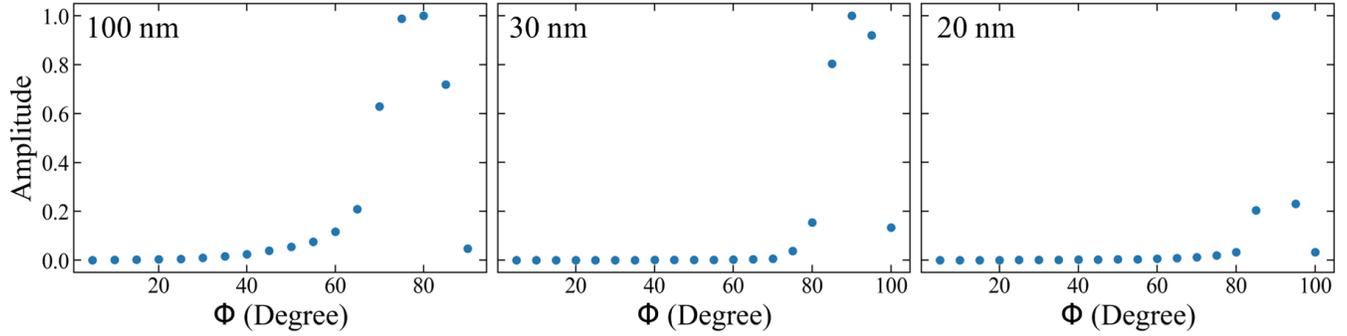

**Figure 8**: *FFT amplitude of auto oscillations as a function of angle of effective field for different constriction widths.*

Auto oscillation amplitude for different constriction widths at different in plane angles of the uniaxial anisotropy. 100 nm has peak amplitude around 75° and the width of the peak is larger than the other constriction widths. Other two narrow constriction have the peak amplitude at 90°. Still due to the extremely dominant demagnetizing effects at 20 nm NC width the amplitude at 85° is considerably lower than that of the 30 nm.

# APPENDIX D: COEFFICIENT OF NON-LINEAR FREQUENCY AT $K_u = 15\ kJ/m^3$

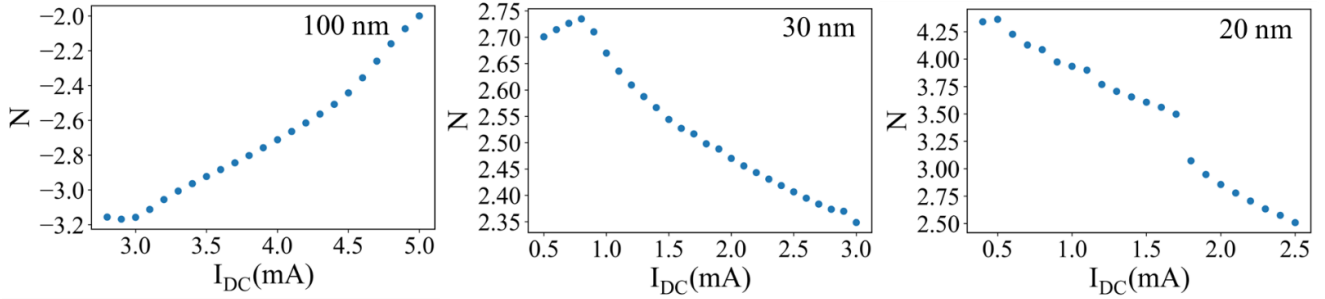

**Figure 9**: *Nonlinear frequency shift for different constriction widths at $K_u = 15\ kJ/m^3$ as a function of input current*

Coefficient of non linear frequency shift (as given in Eq. (6)) is plotted against $I_{DC}$ for 100 nm constriction width at $K_u = 15\ kJ/m^3$. As explained in the main text the shift in frequency from the simulation results approximately matches with the theoretical prediction, where N stays negative for 100 nm and positive for the narrow constrictions. For 30 nm constriction the change in N with current is minute compared to 20 nm and 100 nm constriction which is the reason of poor tunability of frequency with current.

In eqn [7] Slavini *et al.* [14] have assumed an infinite thin film in y-z plane with constant demagnetizing effect along its thickness to get equation (6). The applied field is taken to be in-plane along z direction with a constant magnitude. Resulting excitation is thought of as a standing self-localized wave bullet which is confined in the constriction. Field and in turn the nonlinearity coefficient is assumed to be constant as the current is fixed which differs from our results where both of them are a function of $I_{DC}$. Power of auto oscillations P is also a function of current as amplitude depends on $I_{DC}$.

$$\omega = \omega_o + NP$$

$$\frac{\delta\omega}{\delta I} = N\frac{\delta P}{\delta I} + \frac{\delta N}{\delta I}P$$

Also the frequency of spin wave bullet is single modal while in our case the field free auto oscillation have multiple modes. Thus only a qualitative comparison between the simulation and analytical results is made.

Figures S9 and S10 show the total shift in the frequency of the oscillations with the current in the arbitrary units. For the 100 nm constriction the shift is negative with decrement in the value with increase in $I_{DC}$. The 20 nm constriction show a increasingly positive shift with the current while the 30 nm constriction has a very slow change in shift with the current.

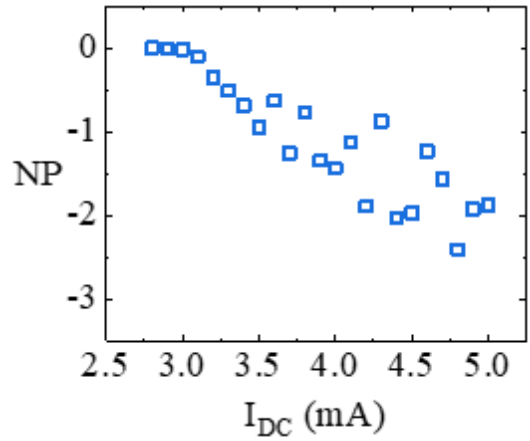

**Figure 10**: *Nonlinear frequency shift × Power for 100 nm constriction width at $K_u = 15\ kJ/m^3$ as a function of input current.*

# APPENDIX E: AUTO-OSCILLATION SPECTRA FOR DIFFERENT CONSTRICTION WIDTHS AS A FUNCTION OF CURRENT

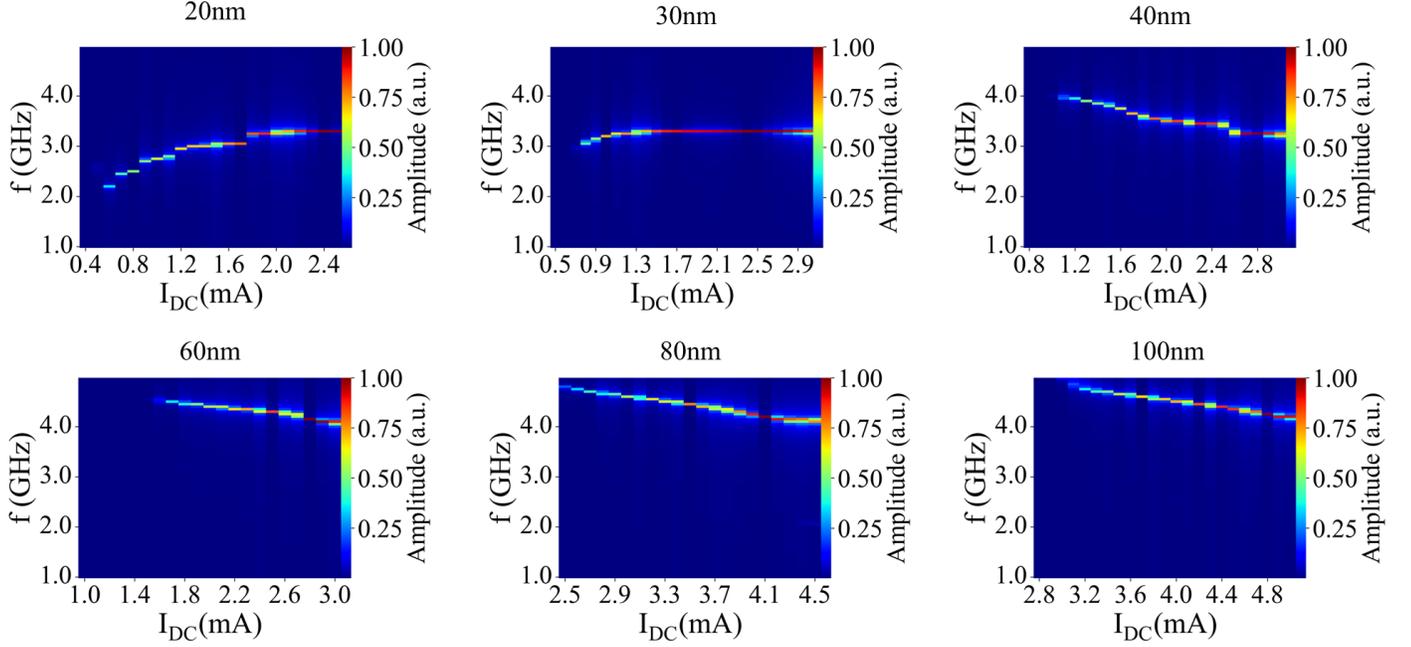

**Figure 11**: *Auto oscillation PSDs at $K_u = 15\ kJ/m^3$ from constriction widths from 20 nm to 100 nm as a function of current are plotted. Colorbars represent the auto oscillation amplitude in arbitrary units.*

Frequency response at 6 different constriction widths (100 nm – 20 nm) is shown. The dynamics till 40 nm constriction width is similar to that of 100 nm, while at 20 nm and 30 nm constriction width the magnetodynamical behaviour is different. Again it is because till 40nm constriction width the $B_{anis}$ contribution to the $B_{int}$ is greater than that of $B_{demag}$ which makes the $B_{int}$ positive. Consequently, the nonlinear frequency shift is negative till 40 nm constriction width. Below the 40 nm constriction width the contribution of $B_{demag}$ towards $B_{int}$ dominates and the magnetodynamic nonlinearity changes sign.

The frequency response with current $\frac{\delta f}{\delta I}$ is positive for narrower constrictions.